\documentclass[12pt,prd,showpacs]{revtex4}%
\usepackage{amssymb}
\usepackage{amsmath}
\usepackage{graphicx}%
\usepackage{amsfonts}
\providecommand{\U}[1]{\protect\rule{.1in}{.1in}}
\begin{document}
\title{Magnetovac Cylinder to Magnetovac Torus }
\author{E.N. Glass}
\affiliation{Department of Physics, University of Michigan, Ann Arbor, MI 48109}
\date{10 August 2006}

\begin{abstract}
A method for mapping known cylindrical magnetovac solutions to solutions in
torus coordinates is developed. Identification of the cylinder ends changes
topology from $R^{1}\times S^{1}$ to $S^{1}\times S^{1}$. An analytic
Einstein-Maxwell solution for a toroidal magnetic field in tori is presented.
The toroidal interior is matched to an asymptotically flat vacuum exterior,
connected by an Israel boundary layer.

\end{abstract}

\maketitle

\section{INTRODUCTION}

Magnetic fields can be found in many astrophysical configurations. Both
poloidal and toroidal fields can model the observed dipolar fields in magnetic
white dwarfs \cite{FW05}. The radius of a nonrotating white dwarf can change
significantly \cite{ST83} when a magnetic field is frozen-in. The equilibrium
configuration of magnetized stars with both poloidal and toroidal magnetic
fields has been studied for some time \cite{IS04}. It is known that toroidal
$\vec{B}$ fields deform neutron stars and it has been shown \cite{Cut02} this
deformation acts as a source of gravitational radiation.

Because of the general interest in astrophysical magnetic fields, this work
provides a method for mapping known cylindrical magnetovac solutions to
magnetovac solutions in nested tori. The cylindrical solution used as an
example has an axially symmetric magnetic field which is mapped to tori with a
toroidal magnetic field. The cylinder solution is unbounded in distance from
the cylinder axis. It is straight forward to map a portion of this solution to
a set of tori. If we conclude the map with an unbounded set of tori, nothing
further is required. But if the torus solution is matched to an exterior
vacuum at a finite boundary, then the metric and extrinsic curvatures, as well
as the magnetic field must also satisfy boundary conditions. The magnetic
field requires a surface current density, which must reside in an Israel
boundary layer \cite{Isr77}. We construct such a layer between the magnetic
tori and the exterior vacuum.

The new solution has four distinct regions. The innermost region is a singular
loop whose time history is a singular cylinder. The next region extends from
the singularity to the vacuum boundary. This second region contains a toroidal
magnetic field in nested tori. At the vacuum boundary there is a third region
of zero thickness with a delta function current density supported by an Israel
layer. The fourth region is an asymptotically flat vacuum.

The next section discusses the map from cylinders to tori and gives details of
the magnetovac cylinder solution. The cylinder solution has a singular z axis.
This is mapped to a singular cylinder in torus coordinates. Section III gives
a new solution for a toroidal magnetic field in torus coordinates. It is shown
that the Rainich conditions for an Einstein-Maxwell solution are satisfied. In
section IV the boundary and extrinsic curvatures are presented. Section V
treats the magnetic field junction conditions, the boundary current, and the
stress-energy content of the Israel layer. In section VI the electromagnetic
energy, Komar mass, and sectional curvature mass are computed. The sectional
curvature mass definition has been extended from spheres to include tori. We
show that the magnetic energy subtracts quasilocally from the sectional
curvature mass. Section VII presents the complete metric and its asymptotic
and inner regions. The vacuum region is asymptotically flat and can be
compactified. We close with a Summary. Details are collected in four
appendices: Appendix A discusses Euclidean torus coordinates. We use
coordinates ($r,\alpha,\beta$); $r$ labels successive tori and ($\alpha,\beta
$) are coordinates on a single torus \cite{Car76}. Appendix B analyzes
geodesics in the plane surrounding the singular region by examining timelike
geodesics paths in an effective potential. Appendix C covers torus null tetrad
quantities. Appendix D lists electromagnetic field equations and the Rainich conditions.

\textbf{Conventions}

Riemann and Ricci sign conventions are $2A_{\nu;[\alpha\beta]}=A_{\mu}%
R_{\ \nu\alpha\beta}^{\mu},$ and $R_{\alpha\beta}=R_{\ \alpha\beta\nu}^{\nu}.$
The metric signature is (+,-,-,-) and the field equations are $G_{\mu\nu
}=-8\pi T_{\mu\nu}$. We use units such that $G=c=1$.

\textbf{Magnetic field terminology}

In cylindrical or spherical coordinate systems, magnetic field $\vec{B}$ has
orthogonal \textit{toroidal }and\textit{\ poloidal }components. A component is
\textit{toroidal} if it is everywhere orthogonal to meridional planes through
the cylinder axis or planes through the spherical poles. If $\vec{B}$ has
axial symmetry, toroidal field lines are circles about the cylinder axis or
spherical circles of longitude. A component lying totally in meridional planes
is \textit{poloidal}, and is orthogonal to the toroidal component.

\section{CYLINDER TO TORUS}

A torus is the only compact, oriented, Euler number zero, 2-dimensional
surface. A cylinder can be made into a torus by joining its ends.

The cylindrical 3-metric is $d\tilde{r}^{2}+\tilde{r}^{2}d\varphi^{2}+dz^{2}
$, and this is mapped to torus coordinates as $dr^{2}+r^{2}d\beta^{2}+\Re
^{2}d\alpha^{2}$. (Details of torus coordinates are given in Appendix A.)
Cylindrical $\tilde{r}\rightarrow$ torus $r$, and cylindrical $\varphi
\rightarrow$ torus $\beta$. For the cylinder ends, $dz=\Re d\alpha$. Since
$-\infty<z<\infty$, and $0\leq\alpha\leq2\pi$, the cylinder is wrapped to a
torus. This identification takes $R^{1}\times S^{1}$ to $S^{1}\times S^{1}$.
The cylinder $z$ axis corresponds to $\Re=r_{0}+r\cos\beta=0$. Because of
torus topology, flat metric $dt^{2}-(dr^{2}+r^{2}d\beta^{2}+\Re^{2}d\alpha
^{2})$ is not the Minkowksi metric.

\subsection*{Cylinder Solution}

The cylindrical metric we consider (Eq.22.11, $m=2$, in \textit{Exact
Solutions} \cite{KSH+03}) is
\begin{equation}
g_{\mu\nu}^{\text{cyl}}dx^{\mu}dx^{\nu}=\tilde{r}^{4}(F/b_{0})^{2}%
(dt^{2}-d\tilde{r}^{2})-\tilde{r}^{-2}(F/b_{0})^{2}d\varphi^{2}-\tilde{r}%
^{4}(b_{0}/F)^{2}dz^{2},\label{cyl-met}%
\end{equation}
with $F(\tilde{r})=1+b_{0}^{2}\tilde{r}^{4}$. This solution satisfies the
Einstein-Maxwell field equations and has an axially symmetric magnetic field
along the $z$ axis. The Kretschmann scalar is%
\[
R^{\alpha\beta\mu\nu}R_{\alpha\beta\mu\nu}=64\frac{b_{0}^{4}(3+12b_{0}%
^{2}\tilde{r}^{4}+62b_{0}^{4}\tilde{r}^{8}-108b_{0}^{6}\tilde{r}^{12}%
+63b_{0}^{8}\tilde{r}^{16})}{\tilde{r}^{12}F^{8}}.
\]
Metric $g^{\text{cyl}}$ has a singularity along the $z$ axis, $\tilde{r}=0$,
where the Kretschmann scalar becomes infinite.

The Ricci tensor has components $t,\tilde{r},\varphi,z$%
\begin{equation}
R_{\ \beta}^{\alpha}=16\frac{b_{0}^{4}}{\tilde{r}^{2}F^{4}}%
\begin{bmatrix}
1 &  &  & \\
& -1 &  & \\
&  & 1 & \\
&  &  & -1
\end{bmatrix}
.\label{ricci-up-dn}%
\end{equation}
The magnetic field is%
\begin{equation}
B_{\mu}dx^{\mu}=4\frac{b_{0}^{2}}{\tilde{r}^{2}F}d\varphi\label{cyl-b-field}%
\end{equation}
with electromagnetic invariants%
\begin{align*}
\frac{1}{2}F_{\mu\nu}F^{\mu\nu}  & =16\frac{b_{0}^{4}}{\tilde{r}^{2}F^{4}}\\
\frac{1}{2}F^{\mu\nu}F_{\mu\nu}^{\ast}  & =0.
\end{align*}

The metric and magnetic field can be mapped, locally, to a different metric
and magnetic field in a frame of nested tori. After the transform, the
resulting metric is $g^{\text{tor}}$ given in Eq.(\ref{tor-met2}).

\section{TORUS SOLUTION}

The toroidal magnetic metric is, with a map of cylinder cordinates
\{$\tilde{r},z,\varphi$\} to torus coordinates \{$r,\alpha,\beta$\}, and with
$\Re=r_{0}+r\cos\beta$ and $\digamma=1+b_{0}^{2}\Re^{4}$,%
\begin{align}
g_{\mu\nu}^{\text{tor}}dx^{\mu}dx^{\nu}  & =\digamma^{2}\Re^{4}dt^{2}-\Re
^{4}(\digamma^{2}\cos^{2}\beta+\digamma^{-2}\sin^{2}\beta)dr^{2}-\digamma
^{2}\Re^{-2}d\alpha^{2}\label{tor-met2}\\
& +\Re^{4}(\digamma^{2}-\digamma^{-2})r\cos\beta\sin\beta~2drd\beta-\Re
^{4}(\digamma^{2}\sin^{2}\beta+\digamma^{-2}\cos^{2}\beta)r^{2}d\beta
^{2}\nonumber
\end{align}
with volume element
\begin{equation}
(r\Re^{5}\digamma^{2})~dt\wedge dr\wedge d\alpha\wedge d\beta.\label{tor-vol}%
\end{equation}
Note that if parameter $p_{0}$ is introduced by $\digamma\rightarrow
\digamma/p_{0}$ in metric $g^{\text{tor}}$, just as parameter $b_{0}$ appears
in the cylinder metric, then the Ricci and Riemann tensors are multiplied by
$p_{0}^{2}$. The limit $p_{0}\rightarrow0$ takes $g^{\text{tor}}$ to flatness
(but not to the Minkowski metric).

The Kretschmann scalar for $g^{\text{tor}}$ is $R_{\alpha\beta\mu\nu}%
R^{\alpha\beta\mu\nu}\sim1/\Re^{12}$. The singularity of $g^{\text{cyl}}$ is
mapped to a singular loop, $\Re=0$, at the center of the torus solution. The
spacetime history of the loop is a cylinder.

Killing symmetries are static $\partial_{t}$ and axial $\partial_{\alpha}$.
Constant time hypersurfaces for static $g^{\text{tor}}$ have unit vector
\begin{equation}
\hat{t}_{\text{tor}}^{\mu}\partial_{\mu}=\digamma^{-1}\Re^{-2}\partial_{t}.
\end{equation}
In torus coordinates, the Ricci tensor has components ($t,r,\alpha,\beta$)%
\begin{equation}
R_{\ \nu}^{\mu}=16\frac{b_{0}^{2}}{\Re^{2}\digamma^{4}}%
\begin{bmatrix}
1 &  &  & \\
& -1 &  & \\
&  & 1 & \\
&  &  & -1
\end{bmatrix}
.\label{ricci-comp}%
\end{equation}
Static metrics have the timelike Killing vector as an eigenvector of the Ricci
tensor. For $g^{\text{tor}}$ we find
\[
R_{\ \nu}^{\mu}\delta_{(t)}^{\nu}=\left(  16\frac{b_{0}^{2}}{\Re^{2}%
\digamma^{4}}\right)  \delta_{(t)}^{\mu}%
\]
with eigenvalue twice the magnetic energy density.

The metric is expanded in a null basis, $g_{\mu\nu}^{\text{tor}}=l_{\mu}%
n_{\nu}+n_{\mu}l_{\nu}-m_{\mu}\bar{m}_{\nu}-\bar{m}_{\mu}m_{\nu}$ (see
Appendix B). The only non-zero Ricci component is
\begin{equation}
\Phi_{11}=8b_{0}^{2}\ \Re^{-2}\digamma^{-4}.\label{phi-11}%
\end{equation}
With Ricci tensor%
\begin{equation}
R_{\mu\nu}^{\text{tor}}=-2\Phi_{11}(l_{\mu}n_{\nu}+n_{\mu}l_{\nu}+m_{\mu}%
\bar{m}_{\nu}+\bar{m}_{\mu}m_{\nu})\label{tor-ricci}%
\end{equation}
the Ricci "square" is%
\[
R_{\text{ }\alpha}^{\mu}R_{\text{ }\nu}^{\alpha}=4(\Phi_{11})^{2}g_{\text{
}\nu}^{\mu},
\]
and with $\Phi_{11}$ in Eq.(\ref{phi-11}) this becomes
\begin{equation}
R^{\mu\nu}R_{\mu\nu}=\left(  32\frac{b_{0}^{2}}{\Re^{2}\digamma^{4}}\right)
^{2}.\label{tor-ricci-sq}%
\end{equation}
Equation (\ref{tor-ricci-sq}) shows the Ricci "square" satisfies Rainich
condition (\ref{rain2b}).

The vacuum limit, $b_{0}\rightarrow0$, is%
\begin{equation}
g_{\mu\nu}^{\text{tor-vac}}dx^{\mu}dx^{\nu}=\Re^{4}dt^{2}-\Re^{4}dr^{2}%
-\Re^{-2}d\alpha^{2}-r^{2}\Re^{4}d\beta^{2}.\label{tor-met2-vacuum}%
\end{equation}
$g^{\text{tor-vac}}$ has zero Ricci tensor and non-zero Riemann tensor.
($R_{\alpha\beta\mu\nu}R^{\alpha\beta\mu\nu})_{\text{tor-vac}}=192/\Re^{12}$.

\subsection*{Electromagnetic Field}

The vector potential for the Maxwell field is
\begin{equation}
A_{\mu}^{\text{tor}}dx^{\mu}=b_{0}\frac{\Re^{4}}{\digamma}(\sin\beta
~dr+r\cos\beta~d\beta),\label{vec-pot2}%
\end{equation}
with
\begin{equation}
F_{\mu\nu}^{\text{tor}}=4b_{0}(\frac{r\Re^{3}}{\digamma^{2}})\left[
\delta_{\mu}^{(r)}\delta_{\nu}^{(\beta)}-\delta_{\mu}^{(\beta)}\delta_{\nu
}^{(r)}\right]  .\label{f-mu-nu}%
\end{equation}
There is no electric field since $F_{\mu\nu}^{\text{tor}}$ has no time
component and there is no current since $\nabla_{\nu}F_{\text{tor}}^{\mu\nu
}=0$. The dual Maxwell field is%
\begin{equation}
F_{\mu\nu}^{\ast}=2b_{0}\left[  \delta_{\mu}^{(\alpha)}\delta_{\nu}%
^{(t)}-\delta_{\mu}^{(t)}\delta_{\nu}^{(\alpha)}\right]  .\label{fstar-mu-nu}%
\end{equation}
The local magnetic field $B_{\mu}^{\text{tor}}=F_{\mu\nu}^{\ast}$ $\hat
{t}_{\text{tor}}^{\nu}$, which satisfies $\nabla_{\mu}B^{\mu}=0$, is
\begin{equation}
B_{\mu}^{\text{tor}}dx^{\mu}=4\frac{b_{0}}{\Re^{2}\digamma}~d\alpha,\text{
\ \ \ }B_{\mu}^{\text{tor}}B_{\text{tor}}^{\mu}=-16\frac{b_{0}^{2}}{\Re
^{2}\digamma^{4}}.\label{b-tor-magnetic}%
\end{equation}
The magnetic field lines are toroidal, i.e. $B_{\text{tor}}^{\mu}$ is
everywhere orthogonal to meridional ($r,\beta$) planes through the
$\hat{\alpha}$ axis. The non-zero Maxwell null tetrad component is%
\begin{equation}
\phi_{1}=i(2\sqrt{2}~b_{0})\Re^{-1}\digamma^{-2}\label{phi-1}%
\end{equation}
The relation $\Phi_{ab}=\phi_{a}\bar{\phi}_{b}$ between the Ricci $\Phi_{ab}$
and the Maxwell $\phi_{a}$ must be satisfied, and such is the case for the
solution given here. This is seen explicitly since $\phi_{1}\bar{\phi}%
_{1}=\Phi_{11}$ given in Eq.(\ref{phi-11}). The Maxwell invariants are%
\begin{align}
I_{1}^{\text{tor}}  & =\frac{1}{2}F^{\mu\nu}F_{\mu\nu}=-(\phi_{1}^{2}%
+\bar{\phi}_{1}^{2})=16\frac{b_{0}^{2}}{\Re^{2}\digamma^{4}}%
.\label{max-invars2}\\
I_{2}^{\text{tor}}  & =\frac{1}{2}F^{\mu\nu}F_{\mu\nu}^{\ast}=0.\nonumber
\end{align}

The Weyl tensor vanishes on a particular torus. From Eqs.(\ref{tor-weyl-comp})
all Weyl components are zero at $\Re_{\text{B}}^{4}=1/b_{0}^{2}$, providing an
effective "magnetic range".

\section{Interior-Exterior metric and Extrinsic curvature}

\subsection{Boundary}

One can match interior metric $g^{\text{tor}}$ to exterior metric
$g^{\text{tor-vac}}$. The boundary surface is located at
\[
r=r_{b}=const.
\]
The metric match is straight forward since $g^{\text{tor-vac}}$ is
$g^{\text{tor}}$ with $b_{0}=0$. The second junction condition requires
extrinsic curvatures to match at $r_{b}$. $\hat{N}^{\mu}$ is the unit normal
at $r_{b} $, $\ \hat{N}_{\mu}\hat{N}^{\mu}=-1$, $\ \hat{N}_{\mu}dx^{\mu}=Ndr,
$%
\begin{equation}
N=\Re^{2}\digamma\lbrack1+b_{0}^{2}\Re^{4}\sin^{2}\beta(1+\digamma
)(1+\digamma^{2})]^{-1/2},\text{ \ }N_{\text{vac}}=\Re^{2}.\label{n-norm}%
\end{equation}
The extrinsic curvature, $K_{\mu\nu}$, is the projected covariant derivative%
\begin{equation}
K_{\mu\nu}^{\text{tor}}=\hat{N}_{\rho;\sigma}\perp_{\ \mu}^{\rho}\perp_{\ \nu
}^{\sigma}.\label{extrin-curv}%
\end{equation}
where $\perp_{\ \mu}^{\rho}=g_{\ \mu}^{\rho}+\hat{N}^{\rho}\hat{N}_{\mu}$
projects into the boundary.

\subsection{Interior}

At $r=r_{b}$, $K_{\mu\nu}^{\text{tor}}$, with $\Re_{b}=r_{0}+r_{b}\cos\beta$
and $\digamma_{b}=1+b_{0}^{2}\Re_{b}^{4}$, has components
\begin{subequations}
\begin{align}
K_{(t)(t)}^{\text{tor}}  & =-\frac{2N_{b}\cos\beta(1+3b_{0}^{2}\Re_{b}^{4}%
)}{\Re_{b}\digamma_{b}},\label{k-in-t-t}\\
K_{(\alpha)(\alpha)}^{\text{tor}}  & =-\frac{N_{b}\cos\beta(1-3b_{0}^{2}%
\Re_{b}^{4})}{\Re_{b}^{7}\digamma_{b}},\label{k-in-alph-alph}\\
K_{(\beta)(\beta)}^{\text{tor}}  & =\frac{N_{b}r_{b}^{2}W_{b}}{\Re_{b}%
\digamma_{b}^{5}}.\label{k-in-beta-beta}%
\end{align}%
\end{subequations}
\[
N_{b}=\Re_{b}^{2}\digamma_{b}[1+b_{0}^{2}\Re_{b}^{4}\sin^{2}\beta
(1+\digamma_{b})(1+\digamma_{b}^{2})]^{-1/2}%
\]%
\begin{align*}
W_{b}  & =3\Re_{b}/r_{b}-2r_{0}/r_{b}+b_{0}^{2}\Re_{b}^{4}(3\Re_{b}%
/r_{b}+2r_{0}/r_{b})-2b_{0}^{4}\Re_{b}^{8}(4\cos^{3}\beta+9\cos\beta
-5r_{0}/r_{b})\\
& -(b_{0}^{6}+b_{0}^{10})\Re_{b}^{4}(10\Re_{b}^{5}/r_{b}-28\Re_{b}^{4}%
\cos\beta\sin^{2}\beta-3\Re_{b}/r_{b}-2r_{0}/r_{b})\\
& +2b_{0}^{8}\Re_{b}^{8}(5\Re_{b}/r_{b}-14\cos\beta\sin^{2}\beta).
\end{align*}
Relevant $g_{\text{tor}}^{\mu\nu}$ components are%
\[
g_{\text{tor}}^{(t)(t)}=\Re^{-4}\digamma^{-2},\text{ \ }g_{\text{tor}%
}^{(\alpha)(\alpha)}=-\Re^{2}\digamma^{-2},\text{ \ }g_{\text{tor}}%
^{(\beta)(\beta)}=-r^{-2}\Re^{-4}(\digamma^{4}\sin^{2}\beta+\cos^{2}%
\beta)^{-1}.
\]
The trace of $K_{\mu\nu}^{\text{tor}}$ is
\begin{equation}
K^{\text{tor}}=-\frac{N_{b}}{\Re_{b}^{5}\digamma_{b}^{3}}[2\cos\beta
(1+3b_{0}^{2}\Re_{b}^{4})-\cos\beta(1-3b_{0}^{2}\Re_{b}^{4})+\frac{W_{b}%
}{\digamma_{b}^{2}(\digamma_{b}^{4}\sin^{2}\beta+\cos^{2}\beta)}%
].\label{tor-trace}%
\end{equation}

\subsection{Exterior}

$K_{\mu\nu}^{\text{tor-vac}}$ has components%
\begin{equation}
K_{(t)(t)}^{\text{tor-vac}}=-2\Re_{b}\cos\beta,\text{ \ }K_{(\alpha)(\alpha
)}^{\text{tor-vac}}=-\cos\beta/\Re_{b}^{5},\text{ \ }K_{(\beta)(\beta
)}^{\text{tor-vac}}=r_{b}\Re_{b}(3\Re_{b}-2r_{0}).\label{ext-vac-comps}%
\end{equation}
Relevant $g_{\text{tor-vac}}^{\mu\nu}$ components are%
\[
g_{\text{tor-vac}}^{(t)(t)}=\Re^{-4},\text{ \ }g_{\text{tor-vac}}%
^{(\alpha)(\alpha)}=-\Re^{2},\text{ \ }g_{\text{tor-vac}}^{(\beta)(\beta
)}=-r^{-2}\Re^{-4}.
\]
The trace of $K_{\mu\nu}^{\text{tor-vac}}$ is%
\begin{equation}
K^{\text{tor-vac}}=-\frac{1}{\Re_{b}^{3}}(\cos\beta+3\Re_{b}/r_{b}%
-2r_{0}/r_{b}).\label{vac-trace}%
\end{equation}

\section{Magnetic field match}

With unit normal $\hat{N}$, the magnetic field at the boundary requires%
\begin{equation}
(\vec{B}_{\text{tor}}\cdot\hat{N})_{b}=(\vec{B}_{\text{vac}}\cdot\hat
{N}_{\text{vac}})_{b}.\label{b-perp}%
\end{equation}
This condition is satisfied since the normal components are zero on both sides
of the boundary. The tangential component must obey%
\begin{equation}
(\vec{B}_{\text{tor}}\times\hat{N})_{b}=-\vec{J}_{\text{s}}\label{b-tan}%
\end{equation}
where $\vec{J}_{\text{s}}$ is a surface current density. The jump in
tangential component (non-zero to zero) requires an Israel surface layer to
support $\vec{J}_{\text{s}}$.

\subsection*{Israel Layer}

If the extrinsic curvatures do not match at the boundary, an Israel boundary
layer is created.\ The stress-energy content of the Israel layer is
constructed from the mismatch in the curvatures. The stress-energy of the
boundary layer is \cite{Poi04}
\begin{equation}
-8\pi S_{\ \nu}^{\mu}=\text{ }[K_{\ \nu}^{\mu}]-[K]\perp_{\ \nu}^{\mu}.
\end{equation}
Here $K=K_{\mu}^{\mu}$, and $[K]:=K^{\text{exterior}}-K^{\text{interior}}$
denotes the jump of $K$ (for any tensorial quantity defined on both sides of
the boundary). The non-zero components of the stress-energy are $S_{(t)}%
^{(t)}$, $S_{(\alpha)}^{(\alpha)}$, and $S_{(\beta)}^{(\beta)}$. We find
\begin{subequations}
\begin{align}
\lbrack S_{(t)}^{(t)}]  & =-\frac{1}{\Re_{b}^{5}\digamma_{b}^{3}}[\Re_{b}%
^{2}\digamma_{b}^{3}(\cos\beta+2r_{0}/r_{b}-3\Re_{b}/r_{b})+\frac{N_{b}W_{b}%
}{\digamma_{b}^{4}\sin^{2}\beta+\cos^{2}\beta}\label{s-t-t}\\
& -N_{b}\cos\beta(1-3b_{0}^{2}\Re_{b}^{4})]\nonumber\\
\lbrack S_{(\alpha)}^{(\alpha)}]  & =\frac{1}{\Re_{b}^{5}\digamma_{b}^{3}}%
[\Re_{b}^{2}\digamma_{b}^{3}(2\cos\beta+3\Re_{b}/r_{b}-2r_{0}/r_{b}%
)-\frac{N_{b}W_{b}}{\digamma_{b}^{4}\sin^{2}\beta+\cos^{2}\beta}%
\label{s-alph-alph}\\
& -2N_{b}\cos\beta(1+3b_{0}^{2}\Re_{b}^{4})]\nonumber\\
\lbrack S_{(\beta)}^{(\beta)}]  & =\frac{\cos\beta}{\Re_{b}^{5}\digamma
_{b}^{3}}\left[  \Re_{b}^{2}\digamma_{b}^{3}-N_{b}(1+9b_{0}^{2}\Re_{b}%
^{4})\right] \label{s-beta-beta}%
\end{align}
When the magnetic field is set to zero, i.e. $b_{0}^{2}\rightarrow0$, all
jumps vanish.

A graph of $S_{(t)}^{(t)}$ for weak magnetic field, with $\digamma_{b}^{2}%
\sim1$, is bell-shaped and peaked at ($\beta$)$~=$~$\pi$. (Fig. 1) With large
magnetic field, $\digamma_{b}^{2}\sim b_{0}^{4}\Re_{b}^{8},$ the graph has two
bell-shaped peaks at ($\beta$)$~=$~$\pi/2,3\pi/2$ falling to zero at ($\beta
$)$~=$~$\pi$. (Fig. 2) The other components have similar graphs. The
stress-energy is peaked at the "top, $\pi/2$" and "bottom, $3\pi/2$" of a
given torus. Since magnetic energy subtracts quasilocally from mass energy,
the stress-energy is greater at top and bottom to maintain the torus shape.
With no magnetic energy, the stress-energy is a maximum at the innermost
point, ($\beta$)$~=$~$\pi$, to balance mass attraction.%
\begin{figure}
[h]
\begin{center}
\includegraphics[
height=2.5633in,
width=3.5362in
]%
{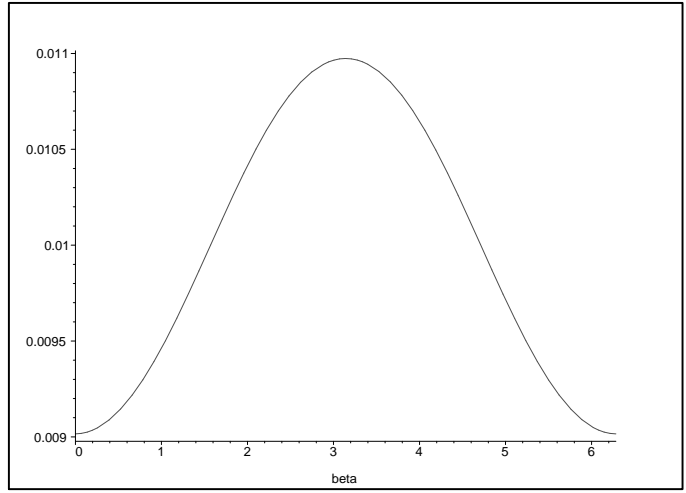}%
\caption{$r_{0}/r_{b}=10$. Weak magnetic field with $F_{b}^{2}\sim1$. Vertical
scale adusted to graph. }%
\label{fig1}%
\end{center}
\end{figure}
\begin{figure}
[h]
\begin{center}
\includegraphics[
height=2.5633in,
width=3.5362in
]%
{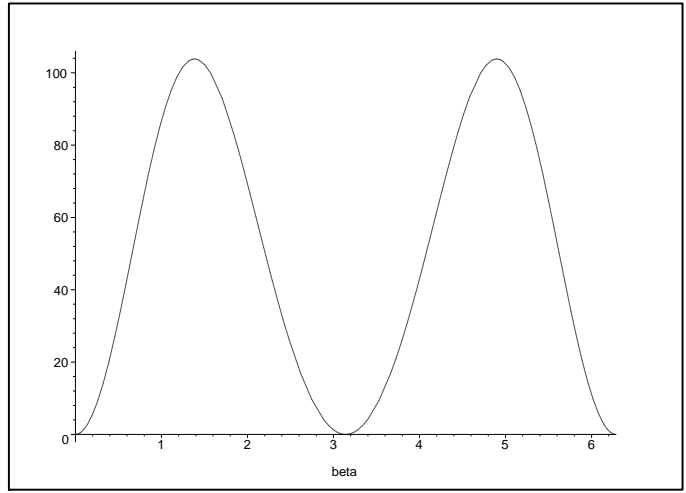}%
\caption{Large magnetic field. $F_{b}^{2}\sim b_{0}^{4}\Re_{b}^{8}$. Vertical
scale adjusted to graph with $b_{0}=1$.}%
\label{fig2}%
\end{center}
\end{figure}

In the outer equatorial plane, with $\cos\beta=1$,
\end{subequations}
\begin{subequations}
\begin{align}
\lbrack S_{(t)}^{(t)}]  & \simeq b_{0}^{2}(1+r_{0}/r_{b})(r_{b}-4r_{0}%
)+O(r_{b}{}^{5}),\label{approx-t-t}\\
\lbrack S_{(\alpha)}^{(\alpha)}]  & \simeq-2b_{0}^{2}(1+r_{0}/r_{b}%
)(r_{b}-2r_{0})+O(r_{b}{}^{2}),\label{approx-alph-alph}\\
\lbrack S_{(\beta)}^{(\beta)}]  & \simeq-b_{0}^{2}(1+r_{0}/r_{b})+O(r_{b}%
{}^{5}).\label{approx-beta-beta}%
\end{align}
Poisson \cite{Poi04} expresses the complete energy-momentum tensor as
\end{subequations}
\[
T_{\mu\nu}=\Theta(l)T_{\mu\nu}^{\text{exterior}}+\Theta(-l)T_{\mu\nu
}^{\text{interior}}+\delta(l)S_{\mu\nu}.
\]
In this work $T_{\mu\nu}^{\text{exterior}}=0$ for $g^{\text{tor-vac}}$ and
$T_{\mu\nu}^{\text{interior}}=T_{\mu\nu}^{\text{tor}}$ for $g^{\text{tor}}$.
$\Theta$ is the Heaviside step function and $\delta(l)$ is a Dirac
distribution of shell width $l$. We have
\[
T_{\mu\nu}=\Theta(r-r_{b})T_{\mu\nu}^{\text{tor}}+\delta(r-r_{b})S_{\mu\nu}.
\]
The conservation equation for $S_{~\nu}^{\mu}$, its covariant divergence,
expresses the jump in $T_{\mu\nu}^{\text{tor}}$ components at the boundary. To
find the surface current density, we write the field tensor at the boundary%
\[
F_{\text{interior}}^{\mu\nu}=\Theta(r-r_{b})F_{\text{tor}}^{\mu\nu}%
\]
where%
\[
F_{\text{tor}}^{\mu\nu}=(\frac{4b_{0}}{r\Re^{5}\digamma^{2}})[\delta
_{(r)}^{\mu}\delta_{(\beta)}^{\nu}-\delta_{(\beta)}^{\mu}\delta_{(r)}^{\nu}].
\]
$\nabla_{\nu}F_{\text{interior}}^{\mu\nu}=4\pi J_{\text{s}}^{\mu}%
=\delta(r-r_{b})F_{\text{tor}}^{\mu(r)}.$ With $\vec{J}_{\text{s}}$ from
Eq.(\ref{b-tan}) the current is%
\begin{equation}
4\pi J_{\text{s}}^{\mu}\partial_{\mu}=\delta(r-r_{b})(\frac{4b_{0}}{r_{b}%
\Re_{b}^{5}\digamma_{b}^{2}})\partial_{\beta}.\label{surf-j}%
\end{equation}
Any one of the current lines is an $\alpha=const$ circle around the torus.
Together, the circles form a surface current on the bounding torus.%
\begin{figure}
[h]
\begin{center}
\includegraphics[
height=2.5633in,
width=3.5362in
]%
{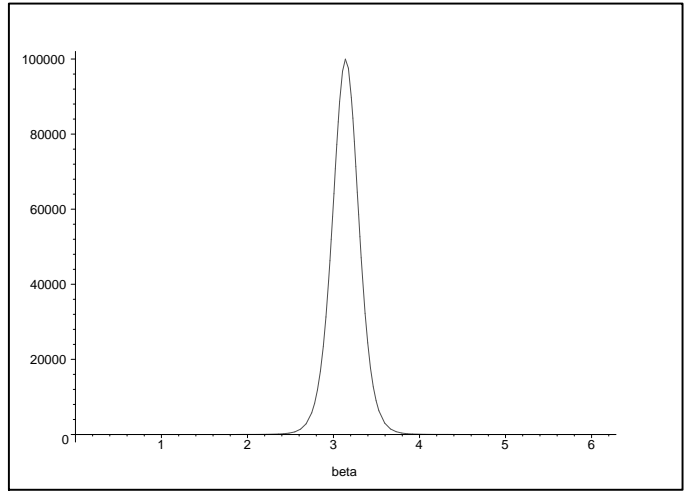}%
\caption{$J_{\text{s}}$ vs $(\beta)$. \ $F_{b}^{2}\sim1,$ $r_{0}/r_{b}=1.1$
\ Vertical scale adjusted to graph.}%
\label{fig3}%
\end{center}
\end{figure}

The graph of one of the current lines, $J_{\text{s}}$ vs ($\beta$), is
bell-shaped and peaked at ($\beta$) $=$ $\pi$. For weak magnetic field, with
$\digamma_{b}^{2}\sim1$, the shape is dominated by $\Re_{b}^{-5}\sim
(r_{0}/r_{b}+\cos\beta)^{-5}$. The curve narrows and the peak rises as
$r_{b}\rightarrow r_{0}$, while the volume containing the magnetic field
shrinks, with a spike as the volume nears zero. With large magnetic field,
$\digamma_{b}^{2}\sim b_{0}^{4}\Re_{b}^{8},$ the current goes as $\Re
_{b}^{-13}$. The curve shape is the same, with smaller magnitude.

\section{Mass and Energy}

\subsection{Magnetic energy}

The electromagnetic energy follows from integrating the magnetic energy
density, $u$, over $t=const$ hypersurfaces within 3-volume $\sqrt{-g}%
d^{3}x=r\Re^{5}\digamma^{2}~d\alpha d\beta dr$%
\begin{align*}
u  & =-\frac{1}{2}B_{\mu}B^{\mu}\\
& =8b_{0}^{2}/(\Re^{2}\digamma^{4}).
\end{align*}
The toroidal energy is%
\[
U_{\text{tor}}(r)=2b_{0}^{2}\int\limits_{r_{0}}^{r}\int\limits_{0}^{2\pi}%
\frac{r^{\prime}(r_{0}+r^{\prime}\cos\beta)^{3}}{[1+b_{0}^{2}(r_{0}+r^{\prime
}\cos\beta)^{4}]^{2}}~d\beta dr^{\prime}.
\]
Termwise integration provides%
\begin{equation}
U_{\text{tor}}(r)=\left[  \frac{b_{0}^{2}r_{0}^{3}}{(1+b_{0}^{2}r_{0}^{4}%
)^{2}}\right]  (r-r_{0})^{2}+O(r-r_{0})^{3}\label{u-tor}%
\end{equation}

\subsection{Komar mass}

The Komar superpotential, with Killing vector $k^{\mu}$, is
\begin{equation}
\mathcal{U}_{\text{komar}}^{\mu\nu}=\sqrt{-g}(\nabla^{\mu}k^{\nu}-\nabla^{\nu
}k^{\mu}),\label{kom-pot}%
\end{equation}
For $k^{\mu}=\delta_{(t)}^{\mu}$ we find%
\[
\mathcal{U}_{\text{komar}}^{\mu\nu}=b_{0}^{2}\frac{\Re^{2}}{\digamma}\left[
r\cos\beta(\delta_{(t)}^{\mu}\delta_{(r)}^{\nu}-\delta_{(t)}^{\nu}\delta
_{(r)}^{\mu})-\sin\beta(\delta_{(t)}^{\mu}\delta_{(\beta)}^{\nu}-\delta
_{(t)}^{\nu}\delta_{(\beta)}^{\mu})\right]  .
\]
The Komar mass is expressed by
\begin{equation}
M(\partial_{t})=\frac{1}{8\pi}\oint\mathcal{U}_{\text{komar}}^{\mu\nu}%
dS_{\mu\nu}.\label{kom-mass}%
\end{equation}
The mass integral, for $t=const$, $r=const$ 2-surfaces with $dS_{\mu\nu
}=2(t_{,\mu}r_{,\nu}-t_{,\nu}r_{,\mu})d\alpha d\beta$, is%
\[
M_{\text{tor}}(r)=4r%
{\displaystyle\int\limits_{0}^{2\pi}}
\cos\beta\left[  \frac{1+3b_{0}^{2}(r_{0}+r\cos\beta)^{4}}{1+b_{0}^{2}%
(r_{0}+r\cos\beta)^{4}}\right]  ~d\beta
\]
Integrating termwise provides the toroidal quasi-local mass%
\begin{equation}
M_{\text{tor}}(r)=\left[  \frac{32\pi b_{0}^{2}r_{0}^{3}}{(1+b_{0}^{2}%
r_{0}^{4})^{2}}\right]  r^{2}+O(r^{3}).\label{tor-mass}%
\end{equation}
The mass expansion is finite since vacuum starts at $r=r_{b}$.

\subsection{Sectional Curvature mass}

Spherical fluids, with 2-metric $r^{2}(d\vartheta^{2}+\sin^{2}\vartheta
d\varphi^{2})$, have a well-defined sectional curvature mass given by
\[
-2m_{\text{sph}}(r)=r^{3}R_{\mu\nu\rho\sigma}\hat{\vartheta}^{\mu}\hat
{\varphi}^{\nu}\hat{\vartheta}^{\rho}\hat{\varphi}^{\sigma}%
\]
where the Riemann tensor is weighted by $r^{3}$ because of spherical geometry.
The Reissner-Nordstr\"{o}m solution (Schwarzschild mass $m_{0}$, charge $q$)
has sectional curvature mass
\begin{equation}
2m_{\text{rn}}(r)=2m_{0}-q^{2}/r.\label{rn-mass}%
\end{equation}
The electrostatic field energy due to charge $q$ has a negative quasilocal contribution.

One can extend this notion to energy in tori. Vectors $\alpha^{\mu}$ and
$\beta^{\mu}$ span $r=const$ tori and are surface-forming, i.e. $\beta^{\nu
}\alpha_{\ ;\nu}^{\mu}-\alpha^{\nu}\beta_{\ ;\nu}^{\mu}=a\alpha^{\mu}%
+b\beta^{\mu}$. We define the torus sectional curvature mass as
\begin{equation}
-2m_{\text{tor}}(r)=[\Re^{6}R_{\mu\nu\rho\sigma}\hat{\alpha}^{\mu}\hat{\beta
}^{\nu}\hat{\alpha}^{\rho}\hat{\beta}^{\sigma}]_{(\beta)=0}\label{m-def}%
\end{equation}
with the Riemann tensor weighted by $\Re^{6}$ because of the torus geometry in
metric (\ref{tor-met2}). Fixing $(\beta)=0$ selects $\Re=r_{0}+r$ for a given
torus. We find
\begin{equation}
m_{\text{tor}}(r)=m_{0}~\frac{1-b_{0}^{2}(r_{0}+r)^{4}}{[1+b_{0}^{2}%
(r_{0}+r)^{4}]^{4}}.\label{sect-curv-m}%
\end{equation}
Parameter $m_{0}$ appears in the Riemann tensor when metric function
$\digamma$ is scaled by $m_{0}$.

We see that the magnetic energy has a negative contribution to $m_{\text{tor}%
}$, just as the Reissner-Nordstr\"{o}m electrostatic energy. In the vacuum
limit, $b_{0}^{2}\rightarrow0$, $m_{\text{tor}}(r)\rightarrow m_{0}$.

\section{COMPLETE METRIC}

Metrics $g^{\text{tor}}$ and $g^{\text{tor-vac}}$ combine to provide%
\begin{equation}
g^{\text{toroidal}}=\left\{
\begin{array}
[c]{c}%
g^{\text{tor}}\text{, \ }r_{0}<r\leq r_{b}\\
g^{\text{tor-vac}}\text{, \ }r_{b}\leq r<\infty
\end{array}
\right. \label{complete-metric}%
\end{equation}

\subsection{Inner region}

The inner torus boundary has coordinates cos$\beta=-1$ and $r=r_{0}$, with
metric functions $\Re=0$, and $\digamma=1$. The toroidal metric goes as
$g^{\text{tor}}\sim\Re^{4}(dt^{2}-dr^{2}-r^{2}d\beta^{2})-(1/\Re)^{2}%
d\alpha^{2}$ and is not well behaved. Since $R^{\mu\nu}R_{\mu\nu}=(32b_{0}%
^{2}/\Re^{2}\digamma^{4})^{2}$ and $R_{\alpha\beta\mu\nu}R^{\alpha\beta\mu\nu
}\sim1/\Re^{12}$, it is clear that $g^{\text{tor}} $ has a singularity at
$r=r_{0},$ cos$\beta=-1$, the $\Re=0$ loop. The spacetime history of the
central singular loop is a singular cylinder. This reflects the singular axis
of the original cylinder metric (\ref{cyl-met}).

\subsection{Asymptotic region}

The vacuum region starts at $r_{b}$. For large distances $r>>r_{0}$, and with
$\mid\cos\beta\mid\simeq1$, metric function $\Re=r_{0}+r\cos\beta$ go as
$\Re\sim r$.

$g^{\text{tor}}$ is matched to a vacuum solution
\[
g_{\mu\nu}^{\text{tor-vac}}dx^{\mu}dx^{\nu}=\Re^{4}(dt^{2}-dr^{2}-\Re
^{-6}d\alpha^{2}-r^{2}d\beta^{2}).
\]
Asymptotically, for $\mid\cos\beta\mid>0$
\begin{equation}
g^{\text{tor-vac}}\sim r^{4}(dt^{2}-dr^{2}-r_{0}^{-6}d\alpha^{2}-r^{2}%
d\beta^{2}).\label{asymp-tor-vac}%
\end{equation}
When $(\beta)$ is near $\pi/2$ or $3\pi/2$ and $\cos\beta\sim\epsilon$ (near
zero)%
\[
g^{\text{tor-vac}}\sim\epsilon^{4}r^{4}(dt^{2}-dr^{2}-r_{0}^{-6}d\alpha
^{2}-r^{2}d\beta^{2}).
\]
There are two directions which do not have a well defined limiting metric:
$(\beta)=$ $\pi/2,$ $3\pi/2$, where the Gaussian curvature of the
($\alpha,\beta$) 2-surface is zero. This difficulty can be removed by choosing
different torus coordinates. Morse and Feshbach \cite{MF53} use torus
coordinates with a hyperbolic radial coordinate. Present coordinates
\{$r,\alpha,\beta$\}$_{\text{tor}}$ map to \{$\mu,\phi,\eta$\}$_{\text{MF}} $.
The analog of $(\beta)$ is angle $(\eta)$. The Gaussian curvature of
($\phi,\eta$) 2-surfaces is%
\[
\mathcal{K}_{\text{MF}}=1-\cos\eta\cosh\mu
\]
and for $(\eta)=\pi/2$, $3\pi/2$, $\mathcal{K}_{\text{MF}}=1$.

With $r^{4}$ as a conformal factor, $g^{\text{tor-vac}}$ can be compactified
and is asymptotically flat. In the compactified diagram, the boundary of the
singular cylinder curves from $I^{-}$ to $I^{+}$ while the null surfaces of
the tori extend to $\mathcal{I}^{+}$. The spacetime of $g^{\text{tor}}$ is
asymptotically flat and foliated by nested tori.

\section{SUMMARY}

We have developed a method for mapping known static cylindrical magnetovac
solutions to static solutions in torus coordinates. The cylinder topology has
been changed from $R^{1}\times S^{1}$ to torus $S^{1}\times S^{1}$ by
identifying the top and bottom cylinder ends. An explicit example has been
presented with a toroidal magnetic field. The example satisfies Rainich
conditions for valid electrovac solutions.

The new solution has three parameters: $r_{0}$ specifies the torus geometry,
$b_{0}^{2}$ the magnetic field and magnetic energy density, and $m_{0}$ is the
sectional curvature mass parameter. The jump in magnetic field is supported by
a surface current density in an Israel layer at the vacuum boundary. The
solution has an asymptotically flat vacuum exterior, and an Einstein-Maxwell
interior with a singular cylinder at the center. By extending sectional
curvature mass to include energy in tori, we have shown that the magnetic
energy subtracts quasilocally from $m_{\text{tor}}$.

There are other exact cylinder solutions, such as the Melvin metric, which
have no singular region but whose map to torus coordinates provide a more
complicated magnetic field. These will be studied in future work, which will
also add rotation to these solutions.

\textbf{ACKNOWLEDGMENT}

I'm indebted to Professor Jean Krisch for reading and commenting on early
versions of this work

\appendix

\section{Torus coordinates in Euclidean 3-space}

A torus can be generated by sweeping a circle, orthogonal to the x-y plane,
about the z-axis. In the x-y plane $r_{0}$ is the distance from the z-axis to
the center of the generating circle, and $r$ is the radius of the generating
circle. For ranges $0<r<\infty,\ 0<\alpha<2\pi,\ 0<\beta<2\pi$, with
$\Re=r_{0}+r\cos\beta$%
\begin{equation}
x=\Re\cos\alpha,\text{ }y=\Re\sin\alpha,\text{ }z=r\sin\beta.\label{tor_coord}%
\end{equation}
Cartesian and toroidal coordinates are related by%
\[
x^{2}+y^{2}+z^{2}=\Re^{2}+r^{2}\sin^{2}\beta.
\]
The torus $r=0$ lies in the $z=0$ plane and has circular radius $r_{0}$.
Looking down the z-axis (about which $\alpha$ has range $0<\alpha<2\pi$) at
the torus, one sees the torus boundaries as two concentric circles. The
$\beta=$ constant surfaces, $0<\beta<2\pi,$ are spheres centered on the
z-axis. In the plane $\beta=\pi/2$, $x^{2}+y^{2}+z^{2}=r_{0}^{2}+r^{2}$.

The Euclidean metric $dx^{2}+dy^{2}+dz^{2}$ with coordinates (\ref{tor_coord})
becomes
\begin{equation}
dr^{2}+\Re^{2}d\alpha^{2}+r^{2}d\beta^{2}.\label{tor-3met}%
\end{equation}
The ($\alpha,\beta$) 2-surface has Gaussian curvature
\begin{equation}
\mathcal{K}=\frac{\cos\beta}{r(r_{0}+r\cos\beta)}.\label{gauss-curv}%
\end{equation}
The angle $\beta$ ranges from $0$ to $2\pi$ over the generating circle, while
$\mathcal{K}$ is negative, zero, and positive.%
\begin{align*}
\mathcal{K}  & <0\text{ \ for }\pi/2<\beta<3\pi/2\\
\mathcal{K}  & =0\text{ \ for }\beta=\pi/2,\ 3\pi/2\\
\mathcal{K}  & >0\text{ \ for }0<\beta<\pi/2,\ 3\pi/2<\beta<2\pi.
\end{align*}
The Euler-Poincar\'{e} characteristic is%
\begin{equation}
\chi=%
{\displaystyle\int\limits_{0}^{2\pi}}
\mathcal{K\ }r(r_{0}+r\cos\beta)d\beta=0\label{euler-char}%
\end{equation}
corresponding to the torus (sphere with one handle).

\section{GEODESICS}

To better understand the region surrounding the central singularity we examine
the geodesic flow. From metric $g^{\text{tor-vac}}$ we have a Lagrangean for
geodesic paths%
\begin{equation}
\mathcal{L}=\frac{1}{2}(\Re^{4}\dot{t}^{2}-\Re^{4}\dot{r}^{2}-\Re^{-2}%
\dot{\alpha}^{2}-r^{2}\Re^{4}\dot{\beta}^{2})\label{lagrangean}%
\end{equation}
where overdots denote $d/ds$. The geodesic equations follow from%
\begin{equation}
\frac{d}{ds}\frac{\partial\mathcal{L}}{\partial\dot{x}^{a}}-\frac
{\partial\mathcal{L}}{\partial x^{a}}=0.\label{lag-eqn}%
\end{equation}
Killing symmetries $\partial_{t}$ and $\partial_{\alpha}$ yield first
integrals
\begin{align*}
\frac{\partial\mathcal{L}}{\partial\dot{t}}  & =E_{0}=\Re^{4}\dot{t},\\
\frac{\partial\mathcal{L}}{\partial\dot{\alpha}}  & =-J_{0}=-\Re^{-2}%
\dot{\alpha}%
\end{align*}
or%
\begin{equation}
\dot{t}=E_{0}/\Re^{4},\text{ \ \ \ }\dot{\alpha}=J_{0}\Re^{2}\label{2-consts}%
\end{equation}
For $x^{a}=(\beta)$, the timelike geodesic equation is%
\begin{equation}
-r^{2}\Re^{4}\ddot{\beta}+2r^{2}\Re^{3}(r\sin\beta)\dot{\beta}^{2}+2\Re
^{3}(r\sin\beta)(\dot{t}^{2}-\dot{r}^{2})+\Re^{3}(r\sin\beta)\dot{\alpha}%
^{2}=0.\label{beta-eqn}%
\end{equation}
If $(\beta)=\pi$ then $\ddot{\beta}=0$. Furthermore, if $\dot{\beta}=0$ and
$\ddot{\beta}=0$, then the orbit remains in the $(\beta)=\pi$ plane. (In order
to examine the singular region, we choose $(\beta)=\pi$ rather than
$(\beta)=0$.) Again, from metric $g^{\text{tor-vac}}$%
\begin{align}
1  & =E_{0}^{2}/\Re^{4}-J_{0}^{2}\Re^{2}-\Re^{4}\dot{r}^{2}-r^{2}\Re^{4}%
\dot{\beta}^{2}\nonumber\\
1  & =E_{0}^{2}/(r_{0}-r)^{4}-J_{0}^{2}(r_{0}-r)^{2}-(r_{0}-r)^{4}\dot{r}%
^{2}\label{rdot-eqn}%
\end{align}

We follow the central force problem of classical mechanics, where $E_{0}%
=\frac{1}{2}m\dot{r}^{2}+V_{\text{eff}}(r)$, and write $\dot{r}^{2}%
+V_{\text{eff}}(r)=const$. Thus, the timelike geodesic paths in the
$(\beta)=\pi$ plane can be described by the effective potential%
\begin{equation}
V_{\text{eff}}(r)=1/(r-r_{0})^{4}+J_{0}^{2}/(r-r_{0})^{2}-E_{0}^{2}%
/(r-r_{0})^{8}.\label{v-eff}%
\end{equation}
%

\begin{figure}
[h]
\begin{center}
\includegraphics[
height=2.5633in,
width=3.5362in
]%
{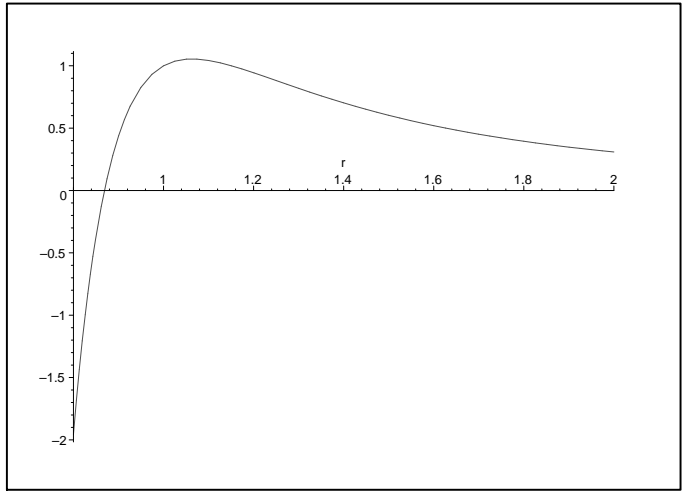}%
\caption{$V_{\text{eff}}$ with $r_{0}=0,$ $J_{0}^{2}=E_{0}^{2}=1$}%
\label{fig4}%
\end{center}
\end{figure}
For $J_{0}^{2}=E_{0}^{2}$, the graph of $V_{\text{eff}}$ has a hard negative
core at $r=r_{0}$. The graph increases exponentially to positive values with a
maximum at $r_{\text{max}},$ and then falls to zero as $1/r^{2}$.
$r_{\text{max}}$ is a positive root of $J_{0}^{2}(r-r_{0})^{6}+2(r-r_{0}%
)^{4}-4E_{0}^{2}=0$.\ At $r=r_{\text{max}}$ there is a closed, unstable,
geodesic path around the singular region.

For $J_{0}^{2}=0$, the shape of the graph is the same, but the height of
$r_{\text{max}}$ is decreased. $r_{\text{max}}=r_{0}+(2E_{0}^{2})^{1/4}$.

For large angular momentum, $J_{0}^{2}>E_{0}^{2}$, the graph has no negative values.

\section{TORUS NULL TETRAD QUANTITIES}

The metric, $g_{\mu\nu}^{\text{tor}}=l_{\mu}n_{\nu}+n_{\mu}l_{\nu}-m_{\mu}%
\bar{m}_{\nu}-\bar{m}_{\mu}m_{\nu}$ with $\Re=r_{0}+r\cos\beta$ and
$\digamma=1+b_{0}^{2}\Re^{4}$, has the null basis
\begin{subequations}
\begin{align}
l_{\mu}dx^{\mu}  & =\frac{\digamma}{\sqrt{2}}(\Re^{2}dt-\Re^{-1}%
d\alpha),\label{el2}\\
n_{\mu}dx^{\mu}  & =\frac{\digamma}{\sqrt{2}}(\Re^{2}dt+\Re^{-1}%
d\alpha),\label{en2}\\
m_{\mu}dx^{\mu}  & =-\frac{\Re^{2}}{\sqrt{2}}[(\digamma\cos\beta
+i\digamma^{-1}\sin\beta)dr-(\digamma\sin\beta-i\digamma^{-1}\cos\beta
)rd\beta],\label{em2}\\
\bar{m}_{\mu}dx^{\mu}  & =-\frac{\Re^{2}}{\sqrt{2}}[(\digamma\cos
\beta-i\digamma^{-1}\sin\beta)dr-(\digamma\sin\beta+i\digamma^{-1}\cos
\beta)rd\beta].\label{em-bar2}%
\end{align}
The spin coefficients are
\end{subequations}
\begin{subequations}
\begin{align}
\epsilon & =\sigma=\lambda=\rho=\mu=\gamma=0\label{zeros2}\\
\kappa & =-\frac{3}{2\sqrt{2}}\frac{1}{\Re^{3}\digamma}\label{kappa2}\\
\nu & =-\kappa\label{nu2}\\
\pi & =\frac{1}{2\sqrt{2}}(\frac{1}{\Re^{3}\digamma^{2}})(1+9b_{0}^{2}\Re
^{4})\label{pie2}\\
\tau & =-\pi\label{tau2}\\
\alpha & =-\frac{1}{\sqrt{2}}(\frac{1}{\Re^{3}\digamma^{2}})(1-9b_{0}^{2}%
\Re^{4})\label{alpha2}\\
\beta & =-\alpha\label{beta2}%
\end{align}
$m_{\mu}$ and $\bar{m}_{\mu}$ are null geodesics, since $m_{\mu;\nu}m^{\nu
}=(\beta-\bar{\alpha})m_{\mu}$. \ The only non-zero Ricci component is
\end{subequations}
\begin{equation}
\Phi_{11}=8\frac{b_{0}^{2}}{\Re^{2}\digamma^{4}}%
\end{equation}
The Weyl tensor null tetrad components are
\begin{subequations}
\label{tor-weyl-comp}%
\begin{align}
\Psi_{0}  & =-(\frac{3}{\Re^{6}\digamma^{3}})(1-b_{0}^{2}\Re^{4})\\
\Psi_{1}  & =\Psi_{3}=0\\
\Psi_{2}  & =-(\frac{1}{\Re^{6}\digamma^{4}})(1-b_{0}^{2}\Re^{4})(1+9b_{0}%
^{2}\Re^{4})\\
\Psi_{4}  & =\Psi_{0}%
\end{align}
which correspond to Petrov type \textbf{I}.

The quadratic Ricci and Weyl invariants are
\end{subequations}
\begin{subequations}
\begin{align}
R^{\mu\nu}R_{\mu\nu}  & =(32b_{0}^{2}/\Re^{2}\digamma^{4})^{2}%
,\label{quad-ricci2}\\
C^{\alpha\beta\mu\nu}C_{\alpha\beta\mu\nu}  & =\frac{192}{\Re^{12}\digamma
^{4}}\frac{(1-b_{0}^{2}\Re^{4})(1+6b_{0}^{2}\Re^{4}+21b_{0}^{4}\Re^{8}%
)}{(1+6b_{0}^{2}\Re^{4}+b_{0}^{4}\Re^{8})}.\label{quad-riem2}%
\end{align}

\section{Relativistic Electromagnetic Field}

The Maxwell field is represented by skew tensor $F_{\mu\nu}$, with dual field
$F_{\mu\nu}^{\ast}~=~\frac{1}{2}\sqrt{-g}\varepsilon_{\mu\nu\alpha\beta}%
$~$F^{\alpha\beta}$. The symmetric energy-momentum is
\end{subequations}
\begin{equation}
T_{\mu\nu}=\frac{1}{4\pi}(F_{\mu\alpha}F_{\text{ }\nu}^{\alpha}+\frac{1}%
{4}g_{\mu\nu}F_{\alpha\beta}F^{\alpha\beta}).\label{t-mu-nu}%
\end{equation}
In a local Lorentz frame, the energy density is%
\begin{equation}
T_{00}=-(E^{2}+B^{2})/8\pi.\label{t-zero-zero}%
\end{equation}
The Einstein field equations provide the trace-free Ricci tensor%
\begin{equation}
R_{\mu\nu}=-(F_{\mu\alpha}F_{\text{ }\nu}^{\alpha}+\frac{1}{4}g_{\mu\nu
}F_{\alpha\beta}F^{\alpha\beta}).\label{ricci-f-f}%
\end{equation}
For static systems there exists a hypersurface orthogonal timelike Killing
vector $\xi^{\mu}\partial_{\mu}=\partial_{t}$. The constant time hypersurfaces
form a family of Lorentz frames with unit normal $\hat{t}^{\nu}=\xi^{\nu}%
/(\xi_{\mu}\xi^{\mu})^{1/2}$. Local electric and magnetic fields are defined
by%
\begin{equation}
E_{\mu}=F_{\mu\nu}\hat{t}^{\nu}=(0,\vec{E}),\text{ \ \ }B_{\mu}=F_{\mu\nu
}^{\ast}\hat{t}^{\nu}=(0,-\vec{B}).
\end{equation}
Maxwell's static equations are%
\begin{equation}
\nabla\cdot\vec{E}=4\pi\rho,\text{ \ \ }\nabla\times\vec{E}=0,\text{
\ \ }\nabla\cdot\vec{B}=0,\text{ \ \ }\nabla\times\vec{B}=4\pi\vec{J}.
\end{equation}
The field invariants are%
\begin{equation}
I_{1}:=\frac{1}{2}F_{\mu\nu}F^{\mu\nu}=B^{2}-E^{2},\text{ \ \ }I_{2}:=\frac
{1}{2}F^{\mu\nu}F_{\mu\nu}^{\ast}=-2E_{\mu}B^{\mu}.
\end{equation}
The Maxwell components on an anti-self dual bivector basis are%
\begin{equation}
\frac{1}{2}(F^{\mu\nu}+iF^{\ast\mu\nu})=\phi_{0}U^{\mu\nu}-\phi_{1}M^{\mu\nu
}+\phi_{2}V^{\mu\nu}.\label{f-bivec}%
\end{equation}
The field invariants are%
\[
I_{1}+iI_{2}=4(\phi_{0}\phi_{2}-\phi_{1}^{2}).
\]

There are three necessary and sufficient Rainich conditions \cite{Wit62} for
$T_{\mu\nu}$ to be an electrovac energy-momentum tensor. \newline1.~$T_{\mu
\nu}$ (and Ricci) must be trace-free.%
\begin{equation}
g^{\mu\nu}T_{\mu\nu}=g^{\mu\nu}R_{\mu\nu}=0.\label{rain1}%
\end{equation}
\newline2.~The "square" of $T_{\mu\nu}$ is a positive multiple of the unit
tensor.
\begin{subequations}
\begin{align}
T_{\text{ }\alpha}^{\mu}T_{\text{ }\nu}^{\alpha}  & =(I_{1}^{2}+I_{2}%
^{2})\delta_{\text{ }\nu}^{\mu}\label{rain2a}\\
\text{or }R_{\text{ }\alpha}^{\mu}R_{\text{ }\nu}^{\alpha}  & =\frac{1}%
{4}(R_{\alpha\beta}R^{\alpha\beta})\delta_{\text{ }\nu}^{\mu}.\label{rain2b}%
\end{align}
\newline3.~Since the electromagnetic energy-momentum density, $T_{00}$, is
negative in a local Minkowski frame, it is necessary for any timelike vector
$t^{\mu} $ that
\end{subequations}
\begin{equation}
T_{\mu\nu}t^{\mu}t^{\nu}\leq0.\label{rain3}%
\end{equation}


\begin{thebibliography}{99}                                                                                               %
\bibitem {FW05}L. Ferrario and D.T. Wickramasinghe, Mon. Not. R. Astron. Soc.
\textbf{356}, 615 (2005).

\bibitem {ST83}S. L. Shapiro and S.A. Teukolsky, \textit{Black Holes, White
Dwarfs, and Neutron Stars}, (Wiley, New York, 1983) Ch. 7.

\bibitem {IS04}K. Ioka and M. Sasaki, Ap. J. \textbf{600}, 296 (2004).

\bibitem {Cut02}C. Cutler, Phys. Rev. D \textbf{66}, 084025 (2002).

\bibitem {Car76}M. P. do Carmo, \textit{Differential Geometry of Curves and
Surfaces}, (Prentice-Hall, New Jersey, 1976) p 157.

\bibitem {Isr77}W. Israel, Phys. Rev. D \textbf{15}, 935 (1977).

\bibitem {KSH+03}\textit{Exact Solutions of Einstein's Field Equations,} Eds.
D. Kramer, H. Stephani, E. Herlt, M. MacCallum and E. Schmutzer, 2nd Ed.
(Cambridge University Press, Cambridge, U.K. 2003).

\bibitem {Poi04}E. Poisson, \textit{A Relativist's Toolkit}, (Cambridge
University Press, 2004) p.89.

\bibitem {Jac99}J.D. Jackson, \textit{Classical Electrodynamics}, (3rd ed.
John Wiley, New York, 1999).

\bibitem {MF53}P.M. Morse and H. Feshbach, \textit{Methods of Theoretical
Physics}, part II, ch 10 (McGraw Hill, New York, 1953).

\bibitem {Wit62}L. Witten, \textquotedblright A Geometric Theory of the
Electromagnetic and Gravitational Fields\textquotedblright\ in
\textit{Gravitation, }Ed. L. Witten, (John Wiley, New York, 1962).
\end{thebibliography}
\end{document}